%% file: main.tex
\documentclass[lettersize,journal]{IEEEtran}
\usepackage{amsmath,amsfonts}
\usepackage{algorithmic}
\usepackage{algorithm}
\usepackage{array}
\usepackage[caption=false,font=normalsize,labelfont=sf,textfont=sf]{subfig}
\usepackage{textcomp}
\usepackage{stfloats}
\usepackage{url}
\usepackage{verbatim}
\usepackage{graphicx}
\usepackage{cite}
\usepackage{tabularx}
\usepackage{booktabs}
\usepackage{ragged2e}
\usepackage{multirow}
\usepackage{caption}
\usepackage{colortbl}
\usepackage{graphicx}
\usepackage{subfig}
\begin{document}

\title{DiffuseTrace: A Transparent and Flexible Watermarking Scheme for Latent Diffusion Model}

\author{Liangqi Lei,
        Keke Gai,~\IEEEmembership{Senior Member,~IEEE,}
        Jing Yu,~\IEEEmembership{Member,~IEEE,}
        Liehuang Zhu,~\IEEEmembership{Senior Member,~IEEE}
}


\IEEEpubidadjcol

\maketitle
\begin{abstract}
\textit{Latent Diffusion Models} (LDMs) enable a wide range of applications but raise ethical concerns regarding illegal utilization.
Adding watermarks to generative model outputs is a vital technique employed for copyright tracking and mitigating potential risks associated with \textit{Artificial Intelligence} (AI)-generated contents. 
However, post-processed watermarking methods are unable to withstand generative watermark attacks and there exists a trade-off between image fidelity and watermark strength. 
Therefore, we propose a novel technique called DiffuseTrace. 
DiffuseTrace does not rely on fine-tuning of the diffusion model components. 
The multi-bit watermark is a embedded into the image space semantically without compromising image quality. 
The watermark component can be utilized as a plug-in in arbitrary diffusion models.
We validate through experiments the effectiveness and flexibility of DiffuseTrace.
Under 8 types of image processing watermark attacks and 3 types of generative watermark attacks, DiffuseTrace maintains watermark detection rate of 99\% and attribution accuracy of over 94\%.
\end{abstract}

\begin{IEEEkeywords}
Latent Diffusion Model, Model Watermarking, Copyright Protection, Security
\end{IEEEkeywords}

\input{introduction}

\input{relatedwork}

\input{background}
\input{problem_formulation}

\input{proposed_scheme}
\input{theoryanalysis}
\input{experiment}

\input{conclusion}

%

\bibliography{ref}
\bibliographystyle{IEEEtran}

\vfill

\end{document}

%% file: introduction.tex
\section{Introduction}
\label{sec:intro}

The advancements made in latent diffusion models \cite{ho2020denoising,song2020score,dhariwal2021diffusion,rombach2022high} have substantially elevated the capability for synthesizing photorealistic content in image generation and have had a profound impact on text-to-image \cite{saharia2022photorealistic,zhang2023adding}, image editing \cite{nichol2021glide,brooks2023instructpix2pix}, inpainting \cite{saharia2022palette,lugmayr2022repaint}, super-resolution \cite{saharia2022image,esser2021taming}, content creation \cite{ramesh2021zero,ramesh2022hierarchical} and video synthesis \cite{blattmann2023align,ho2022imagen}. 
Relevant commercial applications are becoming mainstream creative tools for designers, artists, and the general public.
However, contemporary text-to-image generation models, such as Stable Diffusion and Midjourney, can generate a multitude of novel images, as well as convincing depictions of fabricated events, for malicious purposes. 
Criminals might utilize \textit{Latent Diffusion Models} (LDMs) to produce insulting or offensive images, which could then be disseminated to spread rumors, posing a substantial threat to societal security.
The hazards of deepfakes, impersonation and copyright infringement are also prevalent issues associated with current generative models.



The potential illicit use of text-to-image models has spurred research into embedding watermarks in model outputs. 
Watermarked images contain signals imperceptible to humans but are marked as machine-generated.  
Such watermarks will embed copyright information of the model and identity information about the model's users into the images. 
Extracting watermarks from {\em Artificial Intelligence} (AI)-generated images enables the detection of model copyrights and tracing unauthorized users. 
False and harmful images can be promptly identified and removed from platforms, and unauthorized users of the model can be traced through the extraction of watermark information, thereby mitigating the potential harm caused by AI-generated content.



Existing research on image watermarking mainly focus on post-processing solutions, where the core paradigm revolves around embedding watermarks through minimal image adjustments while emphasizing imperceptibility and precision. 
For instance, the watermarking method implemented in Stable Diffusion \cite{cox2007digital} modifies specific Fourier frequency within the generated image. 
This type of approach inherently encounters a fundamental trade-off between robustness and image quality (e.g., visual fidelity). 
In the context of diffusion model, some work has explored embedding fixed messages into generated images by fine-tuning core components, such as the U-Net architecture or variational autoencoders \cite{ronneberger2015u}. 
However, this type of method has a few limitations. 
First, this approach ‌permits only static information embedding‌, so that it necessitates ‌complete model re-finetuning‌ whenever the embedded content requires modification.
Next, distributed deployment scenarios (e.g., providing customized models to multiple users) ‌require individual fine-tuning for each instance‌, so that computational overhead is caused. 
Finally, iterative model updates ‌may make watermark unreliable due to parameter drift during retraining.
Recent studies \cite{zhao2023invisible} have demonstrated that methods involving the random addition of noise to images to disrupt watermarks, followed by image reconstruction using diffusion models, can effectively remove a significant portion of post-processing watermarking schemes. 
This poses new challenges to the robustness of watermarking techniques.

To address the aforementioned challenges and achieve high extraction accuracy, robustness, and image quality, we propose a new watermarking scheme called DiffuseTrace that fundamentally differs from previous watermarking methods and is compatible with diverse diffusion frameworks. 
Our scheme embeds the watermark into the latent variables of the model during the sampling phase. 
The watermark is embedded at the semantic level prior to image generation, eliminating post-processing of the generated images. 
DiffuseTrace functions as a ‌universal plug-in module‌, requiring neither structural modifications of host models nor fine-tuning procedures.

Taking practical application scenarios into account, we categorize the roles involved in model usage into two types, including model producers and model users. 
Model producers train and own all pre-trained models, i.e., diffusion models, watermark encoders and watermark decoders, and assign specific binary identity information to each user. 
Model producers offer generative model services to users through APIs.
For example, when malicious images resembling model-generated content or images suspected of copyright infringement surface on art platforms, news outlets or other sharing platforms, model producers can trace illegal usage or identify users involved in infringement by extracting watermark information from the generated images.

In the watermark module, we control the watermark distribution by an encoder and dynamically allocate a watermark that approximates the standard normal distribution for each user. 
Since the data distribution and sampling process remain consistent with those of the original model, the generated images achieve transparent watermark embedding while maintaining semantic consistency. 
Human inspection generally cannot distinguish watermarked samples from random samples. 
Watermarks can be decoded by transforming images into latent variables and inversely diffusing them to retrieve the initial latent variables.
Considering the diverse processing stages in the image data flow, as well as the potential bias introduced by the inverse diffusion process of the diffusion model, we employ adversarial training and fine-tune the watermark decoder to improve the robustness of watermark extraction.

The contributions of this work are outlined as follows:
\begin{itemize}

\item The proposed DiffuseTrace is the first scheme that embeds robust multi-bit watermarks, among diffusion watermarking schemes based on initial hidden variables.
DiffuseTrace is embedded at the semantic level of diffusion-model-generated images, eliminating the trade-off between image quality and watermark robustness. 
It exhibits evident advantages over post-processing methods in terms of image quality. 
Compare to the state-of-the-art post-processing watermarking and diffusion model watermarking schemes, DiffuseTrace has a better performance in common image processing and robustness against generative watermark attacks. 
We provide a thorough analysis at the theoretical level regarding the superior watermark robustness of DiffuseTrace.

\item The proposed universal watermark module for latent diffusion models can be seamlessly integrated with different versions of diffusion models. 
The watermark message in DiffuseTrace can be flexibly modified without being affected by fine-tuning or model update iterations. 
We innovatively propose a scheme of latent space encoding and a training mechanism for watermark component.
We conducted extensive comparative experiments on image watermarks based on various schemes.
Our code is open source: https://anonymous.4open.science/r/DiffuseTrace-6DED.

\end{itemize}

The rest of this work is organized as follows. 
Sections \ref{sec:relatedwork} and \ref{sec:background} present reviews of related work and background knowledge, respectively. 
Section \ref{sec:probleformulation} formulates the proposed problem and Section \ref{sec:scheme} provides a detailed description of the proposed scheme. 
We give the theoretical analysis in Section \ref{sec:theory analysis} and experiment evaluations in Section \ref{sec:experiments}.
Section \ref{sec:conclusion} concludes this work.

%% file: relatedwork.tex
\section{Related Work}
\label{sec:relatedwork}

\subsection{Watermark-based Detection for AI-generated Images}

Realistic fake images intensify concerns about the disinformation dissemination. 
\cite{wang2020cnn,frank2020leveraging,tan2023learning} involves extracting temporal, frequency, and texture features from images. 
A feature extraction network is constructed to train a binary classifier to distinguish between AI-generated images and real images.
However, this type of image detection methods causes noticeable performance degradation when applying to diffusion models.
For diffusion model-based AI-generated image detections \cite{wang2023dire}, using a pre-trained diffusion model allows for a more accurate reconstruction of the characteristics of images generated through the diffusion process. 
By reconstructing the diffusion process, differences between real images and images generated by the diffusion model can be detected, thereby enabling the detection of AI-generated images. 



Adding a watermark to the generated images is also a method for identifying AI-generated images.
Traditional image watermarking methods typically involve embedding watermarks into appropriate frequency components of the image, such as {\em Discrete Cosine Transform} (DCT), {\em Discrete Wavelet Transform} (DWT) \cite{al2007combined} or {\em Singular Value Decomposition} (SVD) \cite{liu2019optimized}. 
Deep learning-based approaches, such as HiDDeN \cite{zhu2018hidden}, StegaStamp \cite{tancik2020stegastamp}, have demonstrated competitive results in terms of robustness against various geometric transformations. 
These methods often employ deep learning encoders and extractors to embed and extract watermarks respectively. The aforementioned watermarking methods primarily focus on post-processing existing images. The core idea is to achieve robustness against various attacks while minimizing the impact on the visual quality of the image. Therefore, post-processing methods are confronted with a trade-off between watermark stability, watermark capacity and image quality.

\subsection{Diffusion Model Watermarking}\label{subsec:dmw}
Diffusion model watermarking can be mainly categorized into three types:

\textbf{Watermark embedding during training phase.} 
Watermarks are embedded into the training data. The data is encoded with the watermark during training and a decoder is trained to extract the watermark. During the detection phase, all images generated by diffusion models will carry encoded binary strings. 
Watermark \cite{zhao2023recipe} is a representative approach. Methods of this kind typically have stringent requirements for watermark embedding, involving the incorporation of watermarks into a substantial dataset of images followed by training the entire model.

\textbf{Fine-tuning phase with watermark incorporation.} The main purpose of such watermark embedding methods is to integrate the watermark component into the model component, making it inseparable during distribution.
Watermarks are incorporated into model components during fine-tuning. Methods fine-tune the variational autoencoders to ensure that all generated images carry the watermark, e.g., Stable Signature \cite{fernandez2023stable} and FSwatermark \cite{xiong2023flexible}. 
It is approximate to integrating the watermark into the final generation stage. 

\textbf{Watermark embedding into latent space during inference.} 
Watermarks are added to the latent variable space of the model. 
For example, Tree-ring \cite{wen2023tree} and ZoDiac \cite{zhang2024robust} diffuse inversion and apply frequency domain transformations to latent variables to ensure all generated images carry the watermark. 
DiffuseTrace also falls into this category of methods as the watermark is embedded in the image prior to its generation.

\subsection{Image Watermarking Attack}

The goal of image watermark attacks is to assess the robustness of image detection after practical modifications. These attacks mainly fall into two categories: image processing attacks and deep learning-based attacks. 

\textbf{Image processing attacks.} Common image processing techniques include adding noise, color jitter, image compression, image scaling and Gaussian blur. Image processing or compression methods may utilize frequency-domain or 3D transformation-based approaches including BM3D denoising algorithm 
\cite{dabov2007image}.

\textbf{Deep learning-based attack.} Deep learning-based attack methods, including methods based on variational autoencoders such as \cite{balle2018variational} and \cite{cheng2020learned}
can remove watermarks embedded in images.
In recent research, diffusion based attacks \cite{zhao2023generative} are used to encode the semantic features of images, add noise to remove watermark and regenerate images. Reconstruction models exhibit prominent performance and can eliminate most watermarks injected by most existing methods.

%% file: background.tex
\section{Preliminaries}
\label{sec:background}
\subsection{Diffusion Model-based Image Generation}

Diffusion models progressively transitions the sample x from the true data distribution $p(x)$ to stochastic noise and adeptly reverses this process through iterative denoising of the noisy data \cite{ho2020denoising}. A typical diffusion model framework involves a forward process that progressively diffuses the data distribution $p(x,c)$ towards the noise distribution $p_t(z_t,c)$ for $t \in (0,T]$, where $c$ denotes the conditional context.
The conditional gaussian distribution of the diffusion process is formulated in Eq. (\ref{eq:1}), where $\alpha_t,\sigma_t \in \mathbb{R}^{+}$, $\alpha_t$ and $\sigma_t$ are the strengths of signal and noise respectively decided by a noise scheduler, and $z_t = \alpha_tx + \sigma_t\epsilon$ is the noisy data..
\begin{equation}\label{eq:1}
p_t(z_t|x)=p_t(z_t|\alpha_tx,\sigma_t^2I).
\end{equation}
It has been proved that there exists a denoising process with the same marginal distribution as the forward process \cite{song2020score}. 
The estimation of the only variable derives from Eq. (\ref{eq:2}). 
 \begin{equation}\label{eq:2}
 \nabla _{z_{t}} \log p_t(z_t,c) \approx \frac{\alpha_t x_\theta^{t}(z_t,c)-z_t}{\sigma_t^2}.
 \end{equation}
 Specifically, given a noise-predicting diffusion model parameterized by $\theta$, which is typically structured as a U-Net \cite{ronneberger2015u}, training can be formulated in Eq. (\ref{eq:3}). 
\begin{equation}\label{eq:3}
    \underset{\theta}{min} \quad \mathbb{E}_{x,t,\sigma}||\hat\epsilon_\theta(\alpha_tx+\sigma_t\epsilon,t)-\epsilon||_{2}^{2} , 
\end{equation}
where $t$ refers to the time step; $\epsilon$ is the ground-truth noise; the noise $\epsilon \sim \mathcal{N}(\epsilon|0,I)$ is a standard Gaussian. 
Additionally, LDMs \cite{rombach2022high} streamlines inference processes by incorporating denoising process within the encoded latent space derived from a pre-trained {\em Variational Autoencoder} (VAE) \cite{cemgil2020autoencoding}. 
Diffusion models reconstructs images through the latent state.

During the inference phase, stable diffusion models take both a latent seed and a text prompt as an input. The U-Net progressively removes noise from random latent image representations guided by text embeddings. The noise residual from the U-Net is utilized in conjunction with a scheduler algorithm to generate a denoised latent. 
When synthesizing images, a classifier-free guidance is adopted to enhance the quality of generated images (refer to Eq. (\ref{eq:4})).
\begin{equation}\label{eq:4}
\tilde{\epsilon}_{theta(t,z_t,c)}=w\hat{\epsilon}_{theta(t,z_t,c)}+(w-1)\hat{\epsilon}_{theta(t,z_t,\phi)},
\end{equation}
where the guidance scale $w$ can be modified to regulate the influence of conditional information on the produced images, aiming to strike a balance between quality and diversity. 
$\hat{\epsilon}_{theta(t,z_t,\phi)}$ denotes the unconditional diffusion obtained by an empty prompt. 

\subsection{Diffusion Denoising and Inversion}
The well-trained diffusion model leverages a diverse range of samplers to generate samples from noise and execute denoising procedures. 
A notable denoising method is the {\em Denoising Diffusion Implicit Model} (DDIM) \cite{song2020denoising} which stands out for its efficiency and deterministic output. 
DDIM accomplishes denoising with significantly fewer steps. 
The image $x_0$ will be reproduced with 50 inference steps to the standard 1000-step process. Formally, for each denoising step $t$, DDIM utilizes a learned noise predictor $\epsilon_\theta$ to estimate the noise  $\epsilon_\theta(x_t)$ added to $x_0$, which leads to the estimation of $x_0$ in Eq. (\ref{eq:5}).
\begin{equation}\label{eq:5}
    \hat{x_{0}}=\frac{x_{t}-\sqrt{1-\bar{\alpha}_{t}} \epsilon_{\theta}\left(x_{t}\right)}{\sqrt{\bar{\alpha}_{t}}}.
\end{equation}
The estimated noise $\epsilon_\theta(x_t)$ is recombined with the approximated $\hat{x_{0}}$ to compute $x_{t-1}$, refer to Eq. (\ref{eqa:denoise}).
\begin{equation}
    x_{t-1}=\sqrt{\bar{\alpha}_{t-1}} \hat{x_{0}}+\sqrt{1-\bar{\alpha}_{t-1}} \epsilon_{\theta}\left(x_{t}\right).
    \label{eqa:denoise}
\end{equation} 
DDIM also incorporates an inversion mechanism \cite{dhariwal2021diffusion} to facilitate the reconstruction of the noise representation $x_T$ from an image $x_0$. 
The recovered $x_T$ is mappable to an image approximate to $x_0$. 
Based on the assumption that $x_{t-1} - x_t \approx x_{t+1} - x_t$, the DDIM inversion is formulated in Eq. (\ref{eq:7}).
\begin{equation}
    \hat{x}_{t+1}=\sqrt{\bar{\alpha}_{t+1}} x_{0}+\sqrt{1-\bar{\alpha}_{t+1}} \epsilon_{\theta}\left(x_{t}\right).
    \label{eq:7}
\end{equation}
Essentially, this process follows the forward diffusion process as described in Equation \ref{eqa:denoise}. Diffusion inversion, even in zero-text inversion within conditional diffusion, can still achieve decent accuracy. Meanwhile, the method is applicable to deterministic sampling methods like DPM++ \cite{lu2022dpm}. Our watermarking scheme leverages this property of diffusion inversion.

%% file: problem_formulation.tex
\section{Problem Formulation}
\label{sec:probleformulation}
\begin{figure*}[t]
  \centering
  \includegraphics[width=1.8\columnwidth]{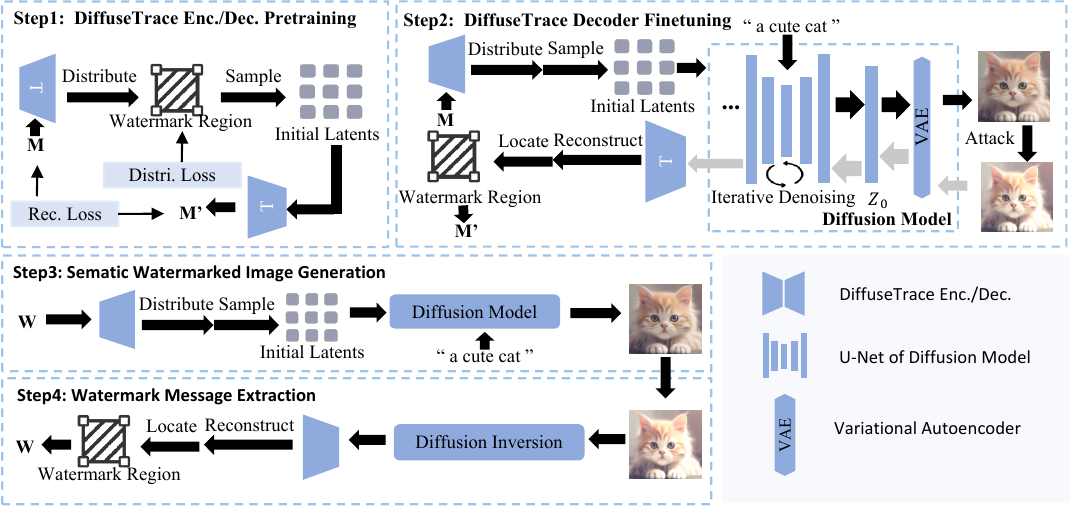}
  
  \caption{Methods of DiffuseTrace. (Step 1) Train the DiffuseTrace Encoder through resampling methods to generate latent variables approximate to a standard normal distribution and jointly train the decoder to decode the information. $M$: Random n-bit messages. (Step 2) Keep the encoder fixed and train the decoder. Randomly select prompts for the diffusion model denoising process to generate images. Decode the images after passing through the attack layer to obtain latent variables and execute diffusion inversion to extract the initial latent variables. Compare the decoded message from the initial latent variables with the initial message to build a reconstruction loss for fine-tuning the decoder. (Step 3) Assign watermark message $w$ and generate initial watermarked latent variables by the encoder to generate images. (Step 4) Extract watermark message after inverting the images and trace the source through statistical testing.}
  \label{fig:method}
\end{figure*}

\subsection{Threat Model}
Two major parties in our work include the defender and adversary. 
A defender is the owner of the generative model. 
Latent diffusion model is deployed as an online service. 
The core objective is to protect the copyright of the model and trace the illegal usage through model outputs. 
The adversary's objective is to disrupt the watermark information in the model output and circumvent the copyright protection and tracing mechanisms of the model.


\textbf{Adversary’s Motivation.}
First, training a latent diffusion model is a costly process, which requires large amount of data, expertise knowledge, and experiments. 
Model parameters are considered proprietary property for business, when considering generative model services a type of online services. 
Adversaries may manipulate images to destroy watermarks and redistributed outputs, causing infringement of intellectual property rights. 
Second, attackers may generate offensive images for malicious purposes, such as fabricating fake news or spreading rumors and remove watermarks from the images to evade tracing.


\textbf{Adversary’s Knowledge and Capability.}
We assume that adversaries can access the victim's latent diffusion model in a black-box manner. 
Attackers can query the victim's latent diffusion model with data samples and obtain corresponding responses.
We categorize adversary background knowledge into two dimensions, including the architecture of the victim's diffusion model and the watermark removal capability. 
For the architecture of the diffusion model, we assume adversaries can access it since such information is typically publicly accessible. 
Regarding watermark removal capability, we assume adversaries can manipulate images via multiple techniques including Gaussian blur, color jittering, and image compression. 
Meanwhile, we consider adversaries who possess the capability to perform state-of-the-art watermark removal attacks by using variational autoencoders and diffusion models.

\subsection{Design Objective}\label{subsec:do}

The objective is to develop a watermark that achieves flexibility, robustness to post-processing operations, generalizability across domains, and preservation of image quality and semantic integrity. 
Such watermarking should exhibit resilience against model fine-tuning or iterative updating procedures.
Three objective properties are given as follows. 


\textbf{Robustness}: 
The watermark is robust against various image processing techniques and state-of-art watermark removal methods. 
Watermarked images are supposed to undergo diverse image processing operations. 
The goal is to ensure that the watermark can be fully recovered even after post-processing interventions. 
The method shall withstand representative watermark removal attacks, e.g. Gaussian noise, color jittering, and Gaussian blur, and defend against the latest watermark attacks based on the state-of-the-art variational autoencoder and diffusion model techniques.


\textbf{Generalizability}: 
Watermark messages can be modified flexibly without retraining or fine-tuning the model.
Given the computational costs associated with embedding fixed information into fine-tuned models, the goal is to achieve adaptable message embedding that maintains compatibility between multiple diffusion model versions. 
The method must demonstrate ‌robustness against model fine-tuning and ‌iterative updates‌ to ensure persistent functionality.


\textbf{Fidelity}: 
This aims at minimizing the impact on the model's watermarked and non-watermarked outputs.
The generated images must maintain strict consistency with the original model's output characteristics, perserving both semantic consistency and image quality. 
The generated watermark samples shall exhibit minimum differences in visual and semantic quality, comparing to normal samples.
Specifically, the watermark is embedded into the initial latent variables at the semantic level without altering semantic consistency and image quality.


\subsection{Overview of the Scheme}


The core idea of DiffuseTrace is to embed the watermark into latent variables. The initial latent variables of the latent space are divided into multiple watermark regions, with each region corresponding to a portion of the watermark information. To ensure both lossless quality and semantic consistency of the image, the embedded watermark should approximate a standard normal distribution and be extractable by the decoder. 

Fig. \ref{fig:method} illustrates the DiffuseTrace framework. Specifically, DiffuseTrace consists of a watermark encoder and watermark decoder. The model owner encodes the initial latent variables through the watermark encoder. The latent variables are then processed through a scheduler guided by prompts and denoised through a U-Net. Afterward, latent variables are decoded by a variational autoencoder into watermarked images. The watermarked images are subjected to an attack layer and decoded back into the latent space. Through diffusion inversion, the original latent variables are restored. The watermark is then extracted from the decoded latent variables through the decoder.

We define the validation for generative image watermarking here. 
The verification scheme for watermarking of generative images is defined by a tuple $\mathsf{Verification}= \langle Trn,Emb,Eva,Vrf \rangle$. 
Training in a watermark encoder and decoder is a fine-tuning or training process that takes training data ($D=\{ x_d,y_d \}$) as input and output for models $Enc[W]$ and $Dec[W]$ by minimizing a given loss L.
\[Trn(D,Arc[\cdot],Enc[\cdot],L)=\{ Enc[W],Dec[W]\}\]

An embedding process for generating watermarked images is an inference process that embeds the signature $Sig$ into latent variables through an encoder and performs inference through the model $Arc[\cdot]$ and prompt $prm$ to output the watermarked image $Pic[Sig]$.
\[Emb(prm,Arc[\cdot],Enc[\cdot],Lat,Sig)=Pic  [Sig] \]

A quality evaluation process is to evaluate whether or not the discrepency of the quality of watermarked images and original images is less than a predefined threshold $\epsilon$.
\[
Eva(Arc[\cdot],M,Lat,\epsilon)= \{True, False\}
\]
\[
|M(Arc[W,Sig], Lat, Enc[\cdot])-M|\leq\epsilon, 
\]
where 
$M(Arc[W,Sig], Lat, Enc[\cdot])$ denotes the image fidelity or semantic consistency tested against a set of watermarked latents. $M$ is the target generation performance.

A verification process checks whether the expected signature $Sig$ of a given generative image can be successfully verified by Decoder $Dec[\cdot]$ when facing image attacks. 
\[
Vrf(Img, Sig,  Atk,Dec[\cdot],\epsilon)=\{ True, False \}
\]

\textbf{Watermark Detection.} DiffuseTrace embeds a k-bit secret message $m \in \{0, 1\}^k$ into the watermark image. 
The watermark detection algorithm includes an extractor to extract the hidden signal $m'$ from the watermarked image, which uses statistical testing to set a threshold $\tau \in \{0,1,2...k\}$ for the extracted bits. 
When the number of matching bits $E(m,m') \geq \tau$, the image is marked as watermarked. 

We establish the hypothesis $H_1$: The image $pic$ is generated by DiffuseTrace against the null hypothesis $H_0$;
Hypothesis $H_0$: The image is not generated by DiffuseTrace. 
Under $H_0$, we assume that the extracted bits $m'_1, m'_2...m'_k$  are independent and identically distributed Bernoulli random variables with a probability of 0.5. $E(m, m')$ follows a binomial distribution $B(k, 0.5)$. 
Type I error (false positive rate (FPR) , $\sigma$) equals the probability of $E(m, m')$ exceeding $\tau$, derived from the binomial cumulative distribution function. 
It has a closed form using the regularized incomplete beta function $I_x(a; b)$.
\begin{equation}
    \begin{aligned}
\epsilon_{1}(\tau) & =\mathbb{P}(E(m, m^{\prime})>\tau \mid H_{0})=\frac{1}{2^{k}} \sum_{i=\tau+1}^{k}(\begin{array}{l}
k \\
i
\end{array}) \\
& =I_{1 / 2}(\tau+1, k-\tau) .
\end{aligned}
\end{equation}
When we reject the null hypothesis $H_0$ with a p-value less than 0.01, we consider the image to be without a watermark. In practice, for a watermark of 48 bits ($k = 48$), at least 34 bits should be extracted to confirm the presence of the watermark. This provides a reasonable balance between detecting genuine watermarks and avoiding false positives.

%% file: proposed_scheme.tex
\section{Watermarking Scheme Design}
\label{sec:scheme}

\subsection{Pre-training Watermark Encoder-Decoder}

We pretrain the encoder-decoder structure for watermark embedding and extraction.
Training the encoder is to construct a unified representation of watermark information and latent variables under a standard Gaussian distribution based on the watermark information and the embedded watermark region.
When binary identity information of the user is inputted into the encoder, it will produce watermark-embedded latent variables that adhere to a standard Gaussian distribution.

\begin{figure}[t]
  \centering
  \includegraphics[width=1\columnwidth]{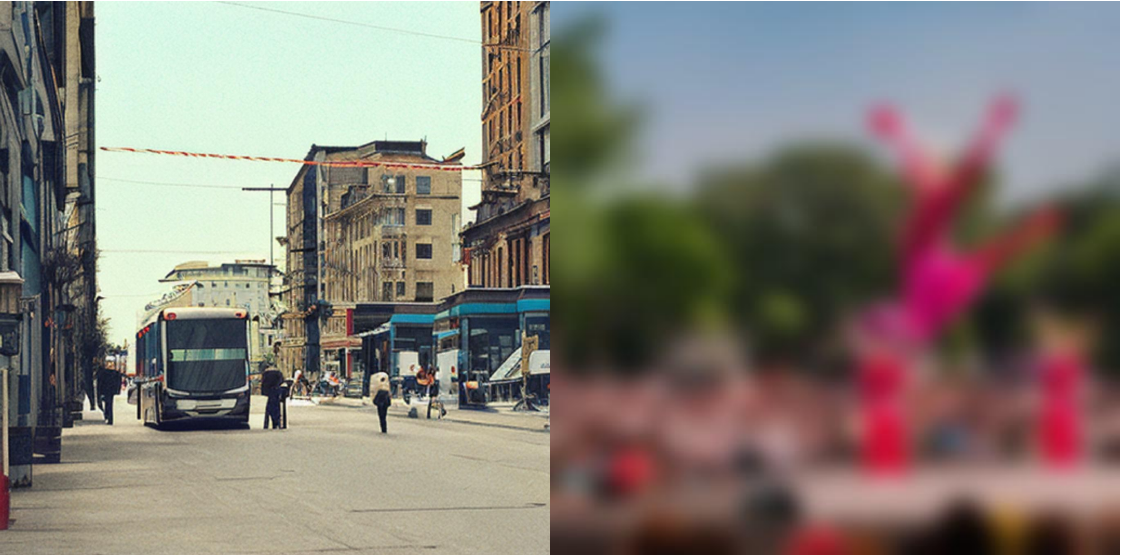}
  \caption{The effect of the deviation of initial latent variables on the image. The left image shows the initial latent variables with a standard normal distribution, while the right image demonstrates a significant deviation. Both the visual quality and semantic quality of the image will experience a substantial decline.}
  \label{pic1}
\end{figure}


We demonstrate that latent variables with watermark confirms to a standard normal distribution. 
To be specific, when the latent generator of LDM samples latent variables $Z$ from noise, the function of the U-Net is will iteratively denoise Gaussian noise matrices within the diffusion cycle guided by text and timesteps.
By subtracting the predicted noise from random Gaussian noise matrices, the random Gaussian noise matrices are eventually transformed into the latent variables of the image.
Since the noise introduced during the training process of the U-Net follows a normal distribution, the initial latent variables of the LDM inference process should ideally approximate standard normal distribution. 
When training a variational autoencoder to encode images into latent variables, one of the training objectives is to approximate the latent variables to adhere roughly to a standard normal distribution. 
The training set for U-net involves repeatedly adding noise from a standard normal distribution to images.
With a sufficient number of iterations, the original images will converge close to a standard normal distribution. 
Therefore, the initial noise is selected to conform to a standard normal distribution during the denoising image generation phase.  
As can be seen in Figure \ref{pic1}, when there is a significant deviation of the initial noise from the standard normal distribution, it may lead to inconsistencies between image quality and semantics. 

Due to the non-differentiability and gradient descent limitations of distributions, the encoder's model architecture employs the reparameterization technique to generate watermark-embedded latent variables. 
Considering the difficulty of explicitly distributing watermark typically regions at a trillion level, we have adopted an implicit partitioning of watermark regions. 
Sampled latent variables are constrained by Kullback-Leibler divergence to approximate a standard normal distribution. 
Each watermark information independently maps to a portion of the probability distribution. 
In addition, the decoder network is the inverse of the encoder. 
The training objective of the decoder is to extract watermark information from the initial latent variables. 
The encoder and decoder are jointly trained to ultimately produce watermark-embedded latent variables that conform to a standard normal distribution. 
The decoder then outputs the corresponding watermark information based on the latent variables.

The reconstruction of watermark refers to maximizing the expected probability distribution of the watermark $w$ given the latent variable. 
Refer to Eq. (\ref{step1_BCE}), the loss function for reconstructing the message $\mathcal{L}_w$ is calculated as the Binary Cross Entropy loss between the original watermark message $m$ and the decoded message $m'$.
\begin{equation}
    \mathcal{L}_w
    =-\sum_{k=0}^{n-1}m_k\log\sigma(m_k')+(1-m_k)\log(1-\sigma(m_k')).
\label{step1_BCE}
\end{equation}

The training samples of the diffusion model are created by gradually involving noises until a standard normal distribution is resembled.
During inference, the message encoder produces initial latent variables that align with the same distribution.
The {\em Kullback-Leibler} (KL) divergence between the initial latent variables and the standard normal distribution is utilized as the loss function. 
The output follows a normal distribution, denoted as $q(z) \sim \mathcal{N}(\mu_1, \sigma_1^2)$, and the standard normal distribution is denoted as $p(z) \sim \mathcal{N}(0, 1)$. 
The distribution loss function is given in Eq. (\ref{eq:dl}).
\begin{equation}\label{eq:dl}
    \mathcal{L}_{dist}=D_{KL}(q(z)||p(z))=\int_xq(x)log\frac{q(x)}{p(x)}dx.
\end{equation}

$\mathcal{L}_w$ ensures the correct decoding of watermark information, while $\mathcal{L}_{dist}$ guarantees the initial distribution of latent variables, thereby ensuring the quality and semantic consistency of the images. 
The encoder and decoder are jointly trained by minimizing the following loss function in Eq.~(\ref{eq:ll}).
\begin{equation}\label{eq:ll}
\mathcal{L}=\lambda_1\mathcal{L}_w+\lambda_2 \mathcal{L}_{dist}.
\end{equation}
$\lambda_1$ and $\lambda_2$ represent the proportion constant parameter.
Therefore, the trained encoder is capable of embedding information into latent variables that approximately adhere to a standard normal distribution, while the decoder is deemed to be the inverse process of the encoder to extract the watermark.

\subsection{Decoder Fine-Tuning}
During the fine-tuning phase of the decoder, the objectives are to adapt to the shift occurring in the watermark region and accurately extract the watermark from the attacked samples. 

\textbf{Watermarking Verification.} Watermark extraction is achieved through diffusion inversion, which approximates the extraction of initial hidden variables from generated images. 
The diffusion inversion \cite{dhariwal2021diffusion} retrieves initial latent variables from images generated by a diffusion model ($x_t$ represents the image at timestep $t$). 
Based on the following assumption \[x_{t-1} - x_t \approx x_{t+1} - x_t,\] diffusion inversion in the {\em Denoising Diffusion Implicit Model} (DDIM) \cite{song2020denoising} is formalized in Eq.~(\ref{eq:ddim}), where $\bar{\alpha}$ is the parameter of the diffusion model, $t$ denotes the denoising timestep, and $\epsilon_\theta(x_t)$ is the estimated noise for timestep $t$.
\begin{equation}\label{eq:ddim}
\hat{x}_{t+1} = \sqrt{\bar{\alpha}_{t+1}} x_{0} + \sqrt{1-\bar{\alpha}_{t+1}} \epsilon_{\theta}(x_{t}).
\end{equation}
The prediction of the image at the current timestep $\hat{x_{0}}$ is defined in Eq.~(\ref{eq:13}).
\begin{equation}\label{eq:13}
    \hat{x_{0}} = \frac{x_{t} - \sqrt{1-\bar{\alpha}_{t}} \epsilon_{\theta}(x_{t})}{\sqrt{\bar{\alpha}_{t}}}.
\end{equation}

A few factors may cause imprecision in decoding throughout the entire watermark embedding-extraction process, which include (i) diffusion inversion process approximates the differences between adjacent step latent variables; (ii) zero-text inversion is applied as the prompt is generally not available during the decoding stage; (iii) potential alterations and manipulations to images may occur through various image processing techniques.
We provide an analysis in Section \ref{subsec:offset}, in which demonstrates that these factors may lead to inevitable deviations in the inferred initial latent variables.

In essence, these processes result in a global shift of the initial latent variables in the semantic space of the images. 
The decoder needs to accurately extract most watermark information, but samples located at the edge of the watermark region exhibit significant inaccuracies. 
To simulate the practical image processing procedures, the attack layer employs an image perturbation technique after generating various images with randomly prompted words. 
The perturbation layer includes randomly adding Gaussian noise, applying Gaussian blur, color jittering and image compression to the images. 
Adversarial training will enhance the robustness of watermark detectors against image processing.

After inverting the images subjected to image processing, we obtain the modified initial latent variables.
We fine-tune the decoder by computing the BCE loss $\mathcal{L}_w$\ref{step1_BCE} between the decoded messages and the original watermark messages as the loss function.

\subsection{Error correction mechanism}
\label{subsec:error correction}
The scheme explicitly delineates the watermark region, while the effects of inversion and image processing can lead to overlap in watermark detection areas during watermark detection due to adversarial training on the decoder.
Thus, it may cause bit errors for samples at the edges of the watermark region where overlap occurs during adversarial training. 
We have provided detailed reasons and explanations (refer to Section \ref{subsec:security analysis}) for employing error correction codes.

\textbf{Recursive Systematic Convolutional (RSC) Codes}: RSC codes provide a systematic approach for encoding and decoding bitstreams, which allows for error correction of data and adaptive recovery of the original message from corrupted data. 
To be specific, given an input bitstream $m$, the RSC encoder transforms it into another bitstream $m+c_1+c_2...c_k$, where each $c_i$ is a bitstream that has the same length as the bitstream $m$ and the symbol $+$ indicates the concatenation of bitstreams. 
A higher encoding ratio can withstand a greater proportion of errors but results in a lengthier encoded bitstream. 


The specific process involves the model owner assigning identity information to the model user, which is then encoded into identity information codes with redundancy using RSC codes.
These identity information codes undergo encoding by the encoder, denoising of latent variables, inversion of latent variables, extraction of watermark information and error correction of the extracted identity information redundancy codes to restore the initial identity information. 
The mechanism of error correction codes combines partial watermark regions into a unified part, correcting the initial latent variables located at the boundaries of watermark detection regions, thereby enhancing the robustness of watermark detection.

%% file: theoryanalysis.tex
\section{Theoretical Analysis}
\label{sec:theory analysis}

\subsection{Unified Representations of Watermark Regions and Latent Variables}
\label{subsec:represent}

Based on the initial requirements, we aim to establish a unified representation for watermark information and latent variable regions. For each watermark $W$, specific distributions of latent variables are distributed. These settings ensure that all images generated by the model can be attributed to the initial distribution of latent variables. 
We set both the diffusion model and the watermark model to share the same latent space. 
For specific parts of this latent space, we sample and extract watermark features based on a probability function $P(z)$. 
Assuming that there are a series of deterministic functions $f(z;\theta)$ parameterized by a vector $\theta$ in some space $\Phi$, where $f: Z \times \Phi \rightarrow \mathbf{X}$. 
When $\theta$ is fixed and $z \sim \mathcal{N}(1,0)$, $f(z; \theta)$ can generate latent variables that conform to a standard Gaussian distribution. 
Adopting this scheme allows watermark distribution regions to be constructed corresponding to specific watermark information. 
The regions coincide with the latent variables of the diffusion model to achieve a unified representation of both. 
The distribution of the watermark conforms to a standard normal distribution. 
This embedding process solely alters the selection of initial latent variables, preserving semantic consistency and image quality.

We aim to optimize $\theta$ such that $z$ can be sampled from $P(z)$ while ensuring the sample closely match the watermark $W$.
The objective of DiffuseTrace is to maximize the probability of each $W$ throughout the entire watermark extraction process, which follows the principle of the maximum likelihood estimation.
When the decoder can reconstruct the watermark from the latent variables, then it can also generate similar samples, and is unlikely to generate dissimilar samples. 
To explain the dependence of $P(z)$ on $W$, we transform $f(z,\theta)$ into $P(W|z;\theta)$. 
The probability density function can be formalized as follows:
\[P(W)=\sum\limits_{z} P(W|z;\theta)P(z).\]
%
The output distribution conforms to a Gaussian distribution after watermark embedding. 
Therefore, $P(W|z;\theta)$ satisfies the following distribution:
\[P(W|z;\theta)=\mathcal{N}(W|f(z;\theta),\sigma^2I).\]
After embedding the watermark, the latent variables have a mean of $f(z; \theta)$ and a covariance equal to the identity matrix $I$ multiplied by the scalar $ \sigma$ which is a hyperparameter.

\subsection{Implicit Allocation of Watermarks}
\label{subsec:distribute}

The principal goal is to partition the standard normal distribution and ensure that each partition can accurately reconstruct the original watermarks. 
For example \cite{doersch2016tutorial}, a 48-bit watermark needs to be divided into over 281 trillion regions, which implies the challenge of manually determining the watermark encoding regions.
Any distribution in $d$ dimensions can be generated by using a set of $d$ variables drawn from a normal distribution and mapped through a function. 
For $P(W)$, within the partitioning of the watermark into a large number of blocks, most sampled $z$ minimally contribute to $P(W)$, since $P(W|z)$ is close to zero for most $z$.
The approximation of the prior distribution can be simplified by using the posterior distribution $q(z|x)$. 
By computing the KL divergence between the posterior and prior distributions, we obtain:
\[D[q(z|w)||p(z|w)]=\mathbb{E}_{z\sim q}[(z|w)-\log p(z|w)].\]
Similar to solving the variational evidence lower bound, we use Bayesian transformation to obtain the lower bound of the reconstructed watermark, shown in Eq.~(\ref{eq:pw}).
\begin{equation}\label{eq:pw}
    \log p(w) \geq \mathbb{E}_{z\sim q}[\log p(w|z)]-D[q(z|w)||p(z)],
\end{equation}
where $\mathbb{E}_{z\sim q}[\log p(w|z)]$ represents maximizing the expected probability distribution of the watermark $w$ given the latent variable $z$, $D[q(z|w)||p(z)]$ is for the approximate posterior distribution of the latent space $z$ to closely resemble the prior distribution.

\subsection{Offset of Watermark Detection Region} 
\label{subsec:offset}

Refer to Eq.~(\ref{eq:7}), diffusion inversion makes the initial latent variables contribute to the generated images. 
The assumption of diffusion inversion approximates $X_{t-1} - X_t$ to $X_{t+1} - X_t$. While unconditional diffusion inversion can lead to accurate results, excessive guidance scale in conditional diffusion amplifies the errors brought by null-text diffusion inversion \cite{mokady2023null}. 

When extracting the semantic embeddings of the images, Existing studies \cite{hong2024exact} have demonstrated conducting a forward pass after each inversion and applying gradient descent can enhance the effectiveness of the inversion process. 
However, considering the computational overhead of the gradient descent process, our proposed scheme accepts the inaccuracy of diffusion inversion under zero text and defines this inaccuracy as the offset to the watermark detection region. 
The purpose of fine-tuning is to learn the offset vector as $p$. 

The watermark encoder trained with maximum likelihood exhibits the following properties, similar samples have more similar latent variables, while dissimilar samples have greater distances in the latent space. 
The distance between the similar latent variables obtained after inversion and the initial latent variables should be less than a certain threshold $\epsilon$ to guarantee the accuracy of detection. 
After the watermark region is segmented, the watermark detection area is offset due to diffusion inversion, the refinement target is defined Eq. (\ref{eqa:loss finetune}).
\begin{equation}
\min_\theta\mathbb{E}_{(x_i,y_i)\sim\mathcal{D}}[\max L(\theta,inv(deno(x_i))+p_i,y_i)],
\label{eqa:loss finetune}
\end{equation}
where $x_i$ denotes specific latent variables,
$y_i$ represents the watermark region to which $x_i$ belongs,
$deno$ represents the process of diffusion denoising, 
$inv$ represents the precise inversion process, 
and $p_i$ denotes the offset of the watermark detection area caused by approximation in Eq. (\ref{eq:7}). 
After fine-tuning the watermark decoder, it should satisfy $p < \epsilon$ to ensure detection accuracy.

Considering that images may undergo various modifications, e.g., image blurring, Gaussian noise, color transformations, such attacks can affect samples on watermark region edges, resulting in a lower-level detection accuracy. 
Essentially, this process does not alter the watermark's region, while it enable repairing the evasion of edge samples. 
Adversarial training can appropriately expand the range of the watermark detection region for various attacks. 
Therefore, the refinement target is formed in Eq. (\ref{eq:17}).
\begin{equation}
\min_\theta\mathbb{E}_{(x_i,y_i)\sim\mathcal{D}}[\max L(\theta,inv(deno(x_i)+\delta)+p_i,y_i)].
\label{eq:17}
\end{equation}
The variable $\delta$ can be expressed as the deviations caused by various attacks on the image, where generated images after such attacks are semantically similar but differs from the original image in semantic space. 
The implementation above implies enhancemnet of the watermark decoder's detection accuracy for edge samples.

\subsection{Security Analysis}
\begin{figure}[t]
  \centering
  \includegraphics[width=1\columnwidth]{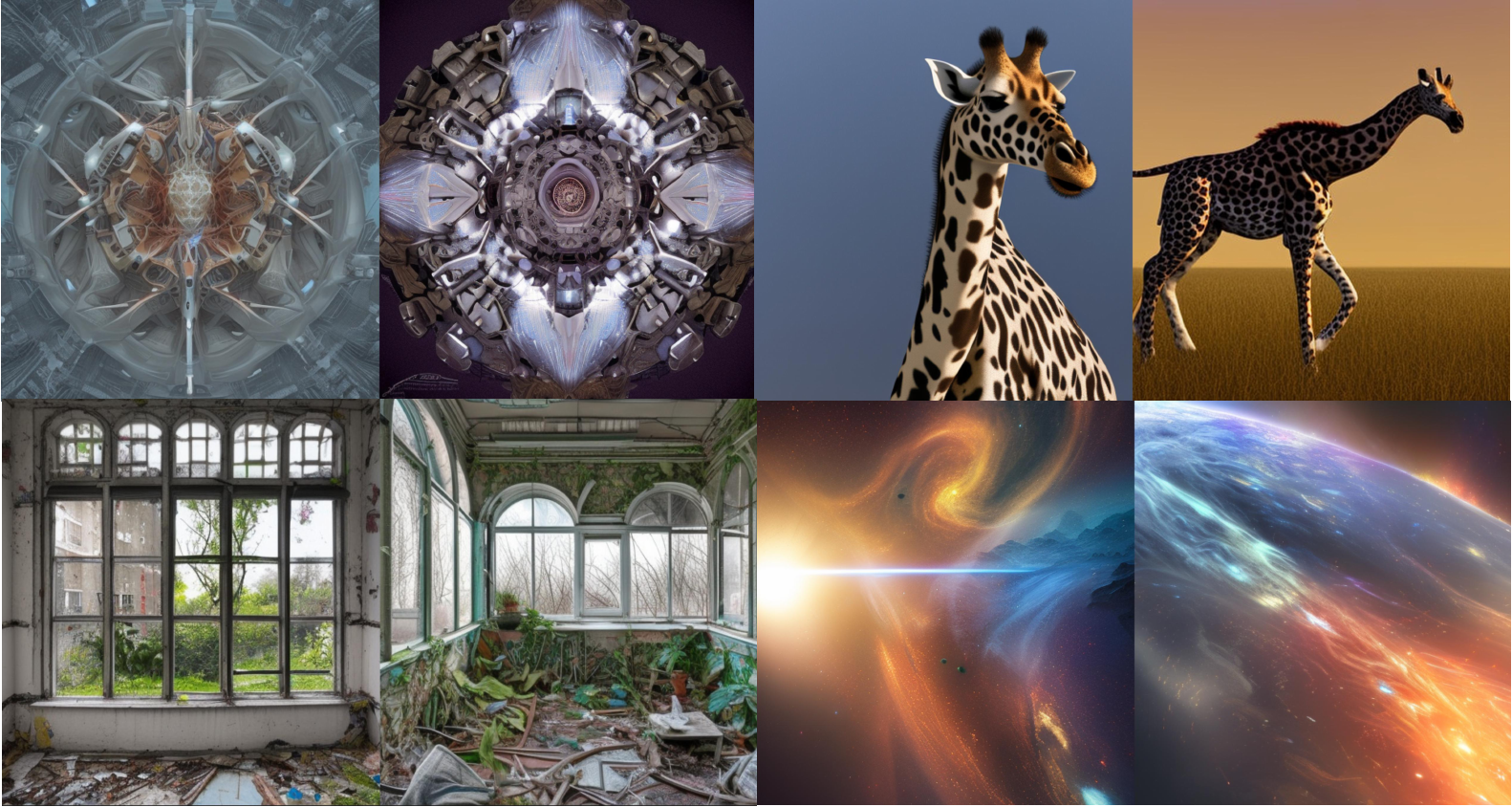}
  \caption{Generated images before (left) and after (right) adding DiffuseTrace. The watermark is embedded at semantic level.}
  \label{pic2}
\end{figure}
\label{subsec:security analysis}

DiffuseTrace divides the watermark region into multiple contiguous areas. 
Assuming the image undergoes image processing resulting in changes compared to the original image, this change is assumed to be $\epsilon_p$ within the latent variable space. 
The initial latent variable corresponding to the original image is $Z_0 \in \mathbf{X}$. 
When there exists $Z_T + \epsilon_p \in \mathbf{X}$, the watermark verification is successful. 
For the initial latent variables close to the center of the watermark region, the distance from the latent variables to other watermark regions is $D \gg \epsilon_p$. 
In this case, watermark verification is straightforward. 
However, for samples at the edges of the watermark region, $Z_T + \epsilon_p \notin \mathbf{X}$. 
In the detection phase, we effectively expanded the detection area for each watermark, considering the outer radius $r$ of each watermark region as part of the region itself. This process can be formalized in Eq. (\ref{eq:21}).
\begin{equation}\label{eq:21}
    detect(z_T+\epsilon_p)=judge(z_0+\epsilon_p \in \mathbf{(X+r)}),
\end{equation}
where the size of $r$ depends on the magnitude of the perturbations in adversarial samples used during adversarial training. 
Expanding the partition of watermark regions actually increases the risk of overlap to some extent between different watermark regions. 

We set the distance $d$ between the two watermark regions. 
Since the encoder remains fixed, the region of the watermark will not be changed. 
However, due to inversion-induced overall shifts and image processing, the detection area post-inversion corresponds to a deviated initial region.
When $r \leq d$, adversarial training enhances the robustness of the watermark to ensure that even edge latent variables can still extract the watermark. 

\textbf{Security Analysis without Attack.} 
When the magnitude of adversarial training $r$ exceeds $d$, it causes the watermark from one edge sample to fall within the detection range of another, leading to bit errors. 
During training, adversarial samples at the boundary regions steer the model in the wrong direction, while correct samples in these regions guide the model back on track. 
As a result, the accuracy of samples in these areas remains above fifty percent but unstable, leading to a fluctuating state.
To correct such errors, we employ error correction codes. 
As discussed in Section \ref{subsec:error correction}, when the error rate of samples in the boundary region is within an acceptable range, error correction codes can restore the original information. 


\textbf{Security Analysis of Image Processing.} 
We consider image manipulation where the same image undergoes a certain offset in the latent space, but within an acceptable range smaller than a certain threshold. 
When the corresponding change in the latent space is less than $d$, adversarial training ensures that both central and marginal latent variables can successfully decode the information. 
Common image manipulations generally keep the image's position in the latent space within an acceptable range, e.g., Gaussian transformations, color jittering, brightness variations, and image compression. 
Therefore, DiffuseTrace effectively defends against such attacks. 

\textbf{VAE-based Attacks and Diffusion-based Attacks.}
VAE-based attacks and Diffusion-based attacks primarily to launch disruption and reconstruction attacks. 
In general, disruption attacks aim at adding noise to the image, while reconstruction involves removing the noise through a diffusion model. 

The reason why such attacks can succeed is that the primary objective of most watermarking schemes is to add minimal watermark noise to the image while still being able to extract the watermark information. These methods often utilize the LPIPS loss \cite{zhang2018unreasonable} or differences in the color channels of the image as the loss function, aiming to minimize the SSIM and PSNR metrics of the final image. This allows reconstruction attacks to exploit this vulnerability by continuously adding noise to gradually degrade the stability of the watermark. Eventually, the reconstruction process generates an image that is indistinguishable from the watermark. While some watermarking schemes, such as Stegastamp, sacrifice image quality and significantly increase adversarial training to enhance their stability, there is no defense against reconstruction attacks when constructive steps become sufficiently numerous. In fact, reconstruction attacks can even produce images that are clearer than the watermark samples.



The watermark based on the initial latent variables primarily operates at the semantic level, allowing for stable watermark extraction as long as there are no significant changes in the image within the latent space. Attacks on watermarks based on diffusion models by adding noise do not alter the original semantic content of the image actually. The initial hidden space positions of the image can still be discerned which makes it resistant to such reconstruction attacks.

%% file: experiment.tex
\begin{table}[t]
\centering
\caption{Image Sematic Quality and Undetectability Evaluation. The table demonstrates the impact of adding semantic watermarks on image quality through two No-inference Metrics, NIQE and PIQE. The semantic consistency before and after adding DiffuseTrace is evaluated through the Clip metric.}
\setlength{\tabcolsep}{0.8mm}{
\begin{tabular}{lccccc}
\toprule
\textbf{Dataset} & \textbf{Method} & \textbf{NIQE$\downarrow$} & \textbf{PIQE$\downarrow$} & \textbf{Clip$\uparrow$} & \textbf{Bit/Detect} \\
\midrule

\multirow{6}{*}{DiffusionDB} 
& No-Watermark & 4.91 & 28.21 & 0.342 & 0.511/0.000 \\
& Tree-ring(rad10) & 5.32 & 30.28 & 0.332 & ------/0.999 \\
& Tree-ring(rad20) & 6.64 & 37.33 & 0.301 & ------/0.999 \\
& DiffuseTrace(16) & 4.22 & 29.08 & 0.344 & 0.999/0.999 \\
& DiffuseTrace(32) & 5.04 & 29.77 & 0.339 & 0.992/0.999 \\
& DiffuseTrace(48) & 4.72 & 28.41 & 0.340 & 0.984/0.999 \\
\midrule

\multirow{6}{*}{\shortstack{MS-COCO\\Prompts}} 
& No-Watermark & 3.85 & 33.28 & 0.335 & 0.504/0.000 \\
& Tree-ring(rad10) & 4.32 & 34.28 & 0.324 & ------/0.999 \\
& Tree-ring(rad20) & 5.64 & 38.33 & 0.291 & ------/0.999 \\
& DiffuseTrace(16) & 4.12 & 33.25 & 0.333 & 0.999/0.999 \\
& DiffuseTrace(32) & 3.81 & 30.21 & 0.326 & 0.994/0.999 \\
& DiffuseTrace(48) & 4.17 & 32.34 & 0.330 & 0.990/0.999 \\
\midrule

\multirow{6}{*}{\shortstack{Diffusion\\Prompts}} 
& No-Watermark & 4.88 & 29.72 & 0.326 & 0.488/0.000 \\
& Tree-ring(rad10) & 5.32 & 30.28 & 0.327 & ------/0.999 \\
& Tree-ring(rad20) & 5.94 & 37.33 & 0.303 & ------/0.999 \\
& DiffuseTrace(16) & 4.93 & 28.42 & 0.358 & 0.999/0.999 \\
& DiffuseTrace(32) & 5.11 & 30.18 & 0.353 & 0.999/0.999 \\
& DiffuseTrace(48) & 4.70 & 26.33 & 0.328 & 0.984/0.999 \\
\bottomrule
\end{tabular}}
\label{tab:sematic evaluation}
\end{table}

\section{Experiments}
\label{sec:experiments}

\subsection{Experiment Configuration}

\textbf{Datasets}. We utilized following datasets:
\begin{itemize}
    \item \textbf{Real Photos}: 500 images were randomly selected from MS-COCO \cite{lin2014microsoft}, which contains over 328K images along with their annotations.
    \item \textbf{Prompts for AI-Generated Images}: 500 prompts were randomly sampled from Diffusion Prompts, a database of approximately 80,000 prompts filtered and extracted from image finders.
    \item \textbf{AI-Generated Images}: 500 images and prompts are randomly chosen from StableDiffusionDB \cite{wang2022diffusiondb}. This dataset contains images generated by Stable Diffusion based on prompts and hyperparameters provided by actual user interactions.
\end{itemize}


\textbf{Watermark Baselines.}
For traditional watermarking schemes, we selected DcTDwt \cite{al2007combined} and DcTDwtSvD \cite{cox2007digital} which is deployed in Stable Diffusion as a watermark with an embedding capacity of 48. For post-processing watermarking schemes based on Encoder-Decoder/GAN structures, we chose RivaGAN 
\cite{zhang2019robust}, Hidden \cite{zhu2018hidden}, SSLWatermark \cite{fernandez2022watermarking} and StegaStamp \cite{tancik2020stegastamp}  with embedding capacities of 32, 48, 48, and 48 respectively. For watermarking schemes based on Variational Autoencoders, we chose Stable Signature \cite{fernandez2023stable} with an embedding capacity of 48. Additionally, for watermarking schemes based on latent variables, we chose Tree-Ring with a watermark radius of 10. Given that tree-ring is a zero-bit watermark scheme, we utilized p-values as the detection metric. The corresponding bit capacity of DiffuseTrace is 48 bits. 

\begin{table*}[htbp]
\caption{Bit Accuracy/Detection Accuracy Under Image Processing. The three settings in the experiment represent the proposed initial method, the method after adversarial training, and the method incorporating adversarial training and error correction codes.}
\centering
\setlength{\tabcolsep}{1.5mm}{
\begin{tabular}{*{10}{c}}
\toprule
\multicolumn{2}{c}{\textbf{Method}} & \textbf{Brightness} & \textbf{Noise} & \textbf{Contrast} & \textbf{Hue} & \textbf{JPEG} & \textbf{Blur} & \textbf{Resize} & \textbf{BM3D}\\
\midrule
\multirow{2}{*}{Traditional Wm.} 
& DwtDct & 0.601/0.168 & 0.801/0.642 & 0.497/0.000 & 0.479/0.000 & 0.488/0.000 & 0.582/0.092 & 0.493/0.000 & 0.498/0.000 \\
& D.Svd & 0.612/0.042 & 0.850/0.999 & 0.718/0.118 & 0.485/0.000 & 0.498/0.000 & 0.989/0.999 & 0.506/0.000 & 0.632/0.084 \\
\midrule
\multirow{4}{*}{Enc.-Dec. Wm.}
& RivaGan & 0.975/0.999 & 0.960/0.994 & 0.832/0.992 & 0.984/0.999 & 0.773/0.801 & 0.867/0.924 & 0.504/0.000 & 0.858/0.873 \\
& Hidden & 0.964/0.999 & 0.971/0.994 & 0.979/0.999 & 0.992/0.999 & 0.849/0.823 & 0.816/0.852 & 0.825/0.873 & 0.626/0.168 \\
& S.Stamp & 0.937/0.999 & 0.979/0.999 & 0.972/0.999 & 0.995/0.999 & 0.952/0.999 & 0.981/0.999 & 0.972/0.999 & 0.980/0.999 \\
& SSLWm. & 0.927/0.999 & 0.627/0.124 & 0.975/0.999 & 0.942/0.997 & 0.547/0.000 & 0.997/0.999 & 0.844/0.901 & 0.620/0.224 \\
\midrule
\multirow{1}{*}{VAE-Based Wm.}
& S.Signa & 0.971/0.999 & 0.976/0.999 & 0.965/0.999 & 0.954/0.994 & 0.852/0.961 & 0.781/0.822 & 0.563/0.141 & 0.744/0.783 \\
\midrule
\multirow{4}{*}{Latent-Based Wm.} 
& Tree-ring & ------/0.999 & ------/0.984 & ------/0.999 & ------/0.999 & ------/0.999 & ------/0.988 & ------/0.999 & ------/0.999 \\
& Ours(ori.) & 0.913/0.999 & 0.915/0.999 & 0.921/0.999 & 0.952/0.999 & 0.891/0.981 & 0.947/0.999 & 0.856/0.950 & 0.881/0.931 \\
& Ours(adv.) & 0.942/0.999 & 0.928/0.999 & 0.959/0.999 & 0.982/0.999 & 0.912/0.999 & 0.966/0.999 & 0.922/0.999 & 0.902/0.999 \\
& Ours(adv.+ecrt.) & 0.949/0.999 & 0.976/0.999 & 0.979/0.999 & 0.988/0.999 & 0.942/0.999 & 0.981/0.999 & 0.940/0.999 & 0.951/0.999 \\
\bottomrule
\end{tabular}}
\label{tab:imageprocess}
\end{table*}

\begin{figure}[t]
  \centering
  \includegraphics[width=0.8\columnwidth]{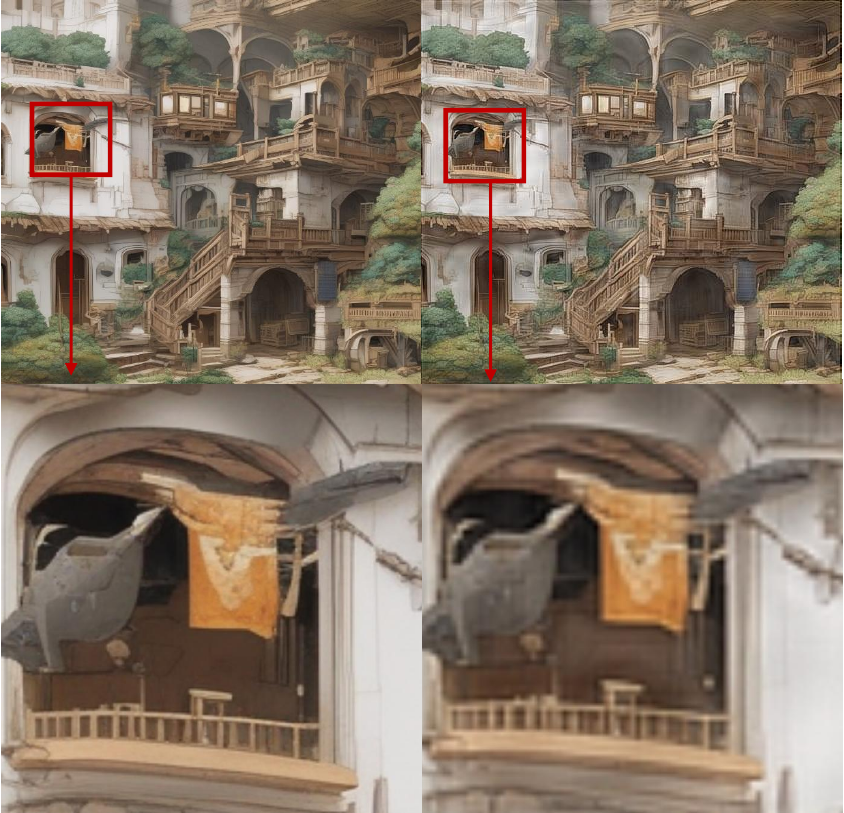}
  \caption{Image quality comparison. The left image was generated using DiffuseTrace (48bit), while the right image was generated using Stegastamp (48bit). It can be observed that post-processing watermark methods will degrade the quality of the image.}
\end{figure}

\textbf{Attack Baselines.} 
To thoroughly evaluate the robustness of DiffuseTrace, we test it against a comprehensive set of baseline attacks that represent common image processing, VAE-based attack and diffusion-based attack. Specially, The set of attacks employed in our testing are listed in Table \ref{tab:attacks}.

\begin{table}[h]
\centering
\caption{Attack Methods for Watermark Testing}
\label{tab:attacks}
\begin{tabular}{ll}
\toprule
\textbf{Method} & \textbf{Description and Parameters} \\
\midrule
Brightness & Global brightness adjustment ($\alpha=2.0$)\\
Noise & Additive Gaussian noise ($\sigma=0.05$) \\
Contrast & Global contrast enhancement ($\gamma=2.0$) \\
Hue & Color hue shift ($\Delta H=0.25$) \\
JPEG & Standard JPEG compression ($q=50$) \\
Blur & Gaussian blur (kernel size $k=7\times7$) \\
Resize & Bilinear image resizing (scale factor $s=0.3$)\\
BM3D & Block-matching 3D denoising (strength $\lambda=30$) \\
\multirow{3}{*}{VAE-based} & Two variational autoencoder compression models: \\
& \quad $\bullet$ BMSHJ2018 \cite{balle2018variational} (quality factor $\lambda=3$) \\
& \quad $\bullet$ Cheng2020 \cite{cheng2020learned} (quality factor $\lambda=3$) \\
Diffusion-based & Diffusion-based image regeneration \cite{zhao2023generative} ($T=40$)\\
\bottomrule
\end{tabular}
\end{table}


\textbf{Evaluation Metrics.} 
The two main objectives of incorporating watermarks are copyright protection and user tracing. Therefore, we utilize $pvalue$ as the standard for copyright tracing and utilize bit accuracy rate as the standard for user tracing. We set a decision threshold to reject the null hypothesis for $p < 0.01$, requiring detection of the corresponding method-corrected 24/32 and 34/48 bits. Otherwise, the image is deemed to be without a watermark.

\textbf{Semantic consistency.} Since our images have watermarks added before image generation, the watermark is reflected at the semantic level. Therefore, we choose the CLIP score metric \cite{Radford2021LearningTV} to evaluate the semantic consistency between generated images and prompt words. The reference metric will be used to evaluate the semantic quality difference between the generated images with and without watermark embedding in DiffuaeTrace in order to evaluate the fidelity of watermark embedding

\textbf{Image Quality.} We evaluate the quality of an image through two no-reference metrics, the Natural Image Quality Evaluator (NIQE) score \cite{mittal2012making} and the Perceptual Image Quality Evaluator (PIQE) score \cite{venkatanath2015blind}. The reference indicators will be used to compare the quality of images without watermarks with images with watermarks embedded in order to evaluate the loss of watermark embedding on the image and the invisibility of the watermark.


\begin{figure*}[h]
  \centering
  \captionsetup[subfloat]{farskip=0pt,captionskip=0pt} 
  \subfloat[Brightness Attack]{\includegraphics[width=0.66\columnwidth]{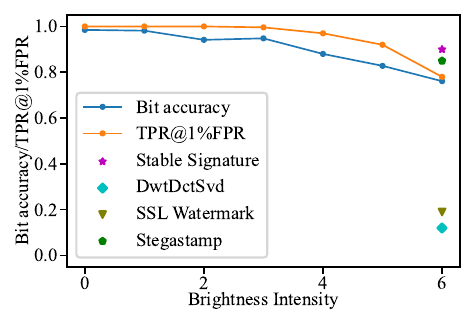}} \hspace{0.01\textwidth}
  \subfloat[Noise Attack]{\includegraphics[width=0.66\columnwidth]{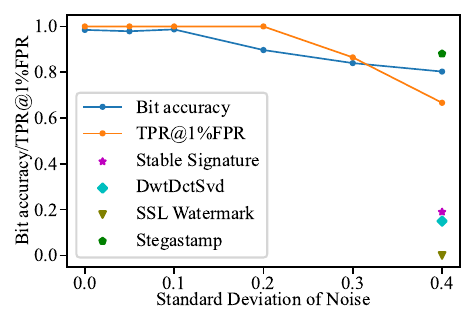}} \hspace{0.01\textwidth}
  \subfloat[Contrast Attack]{\includegraphics[width=0.66\columnwidth]{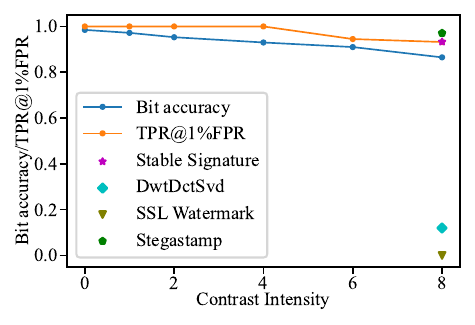}} \\
  \subfloat[JPEG Attack]{\includegraphics[width=0.66\columnwidth]{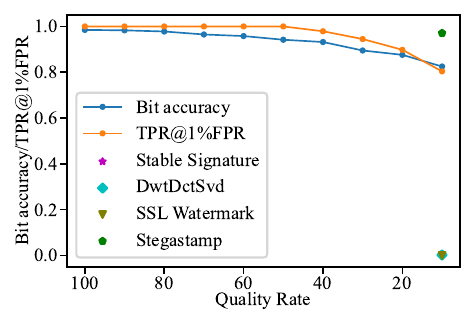}} \hspace{0.01\textwidth}
  \subfloat[Blur Attack]{\includegraphics[width=0.66\columnwidth]{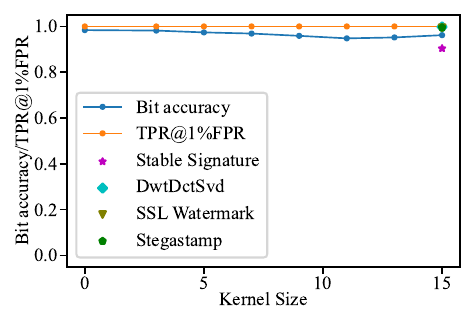}} \hspace{0.01\textwidth}
  \subfloat[Resize Attack]{\includegraphics[width=0.66\columnwidth]{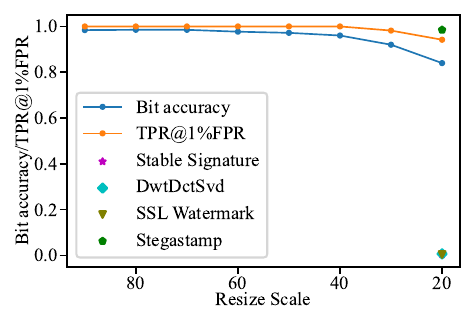}} \\
  \subfloat[BM3D Attack]{\includegraphics[width=0.66\columnwidth]{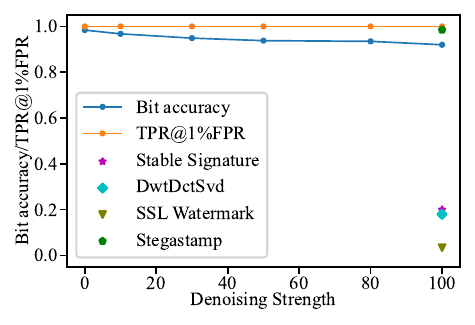}} \hspace{0.01\textwidth}
  \subfloat[VAE-based Attack (Cheng 20)]{\includegraphics[width=0.66\columnwidth]{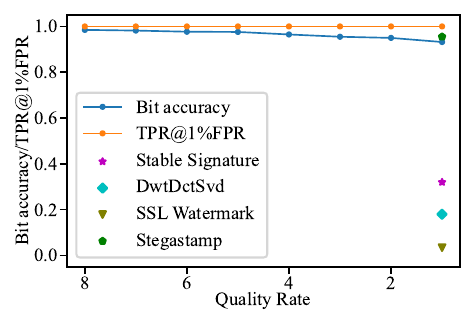}} \hspace{0.01\textwidth}
  \subfloat[Diffusion-based Attack (Zhao 23)]{\includegraphics[width=0.66\columnwidth]{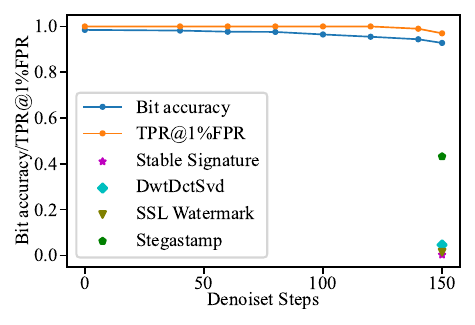}}
  \caption{Performance comparisons of DiffuseTrace in response to various attacks of different intensities, measured by bit accuracy and TPR@0.01FPR. 
  We compare TPR@0.01FPR of traditional watermarks including DwtDctSvd, SSL Watermark, and Stable Signature under corresponding attacks. 
  The watermark capacity for each scheme is 48 bits.}
  \label{deta}
\end{figure*}

\subsection{Sematic and image quality evaluation} 
From Table \ref{tab:sematic evaluation}, we evaluated the impact of embedding the watermark on image quality and semantic consistency before and after adding DiffuseTrace. The experiment utilized the stable-diffusion-2-1-base model \cite{Rombach_2022_CVPR} with 25 inference steps at the guidance scale of 5. Results indicate no significant differences in NIQE and PIQE quality metrics across different watermark bits. Additionally, semantic alignment of generated images, as assessed by Clip scores, remains similar to the original model. This suggests that DiffuseTrace does not rely on the trade-off between image quality and watermark robustness typical of post-processing watermarks. Since images are generated entirely through correct sampling processes and variables are properly distributed without subsequent modifications to the images, the DiffuseTrace approach exhibits significant advantages in terms of image quality compared to post-processing solutions. As can be seen from Figure \ref{pic2}, the Diffusetrace watermark operates at the semantic level.
Compared to other methods that embed watermarks in the latent space, the SSL watermark remains a post-processing solution. Tree-ring alters the distribution of initial latent variables by embedding the watermark in the frequency domain of the latent space through Fourier transformation. However, U-net fails to recover from the losses incurred during this process. Consequently, as the watermark radius increases, the generated images suffer significant losses in both quality and semantic consistency.

\subsection{Robustness evalution against image processing} From Table \ref{tab:imageprocess}, we evaluated the robustness of the watermark against various image processing attacks. It could be observed from the graph that DiffuseTrace demonstrates stability and roubustness when faced with common image processing attacks. DiffuseTrace achieve watermark detection rate close to 100 percent and average bit accuracy above 90 percent under common attack. Compared to post-processing watermarking schemes and VAE-based watermarking schemes, DiffuseTrace demonstrates excellent stability when faced with significant image compression and resizing. Stegastamp
 remain highly robust in the comparison, since it sacrifices image quality. Stable Signature watermark specially for diffusion model remain stable under most attacks, but is vulnerable to denoise algorithm and processing to high extent. Furthermore, We conducted detailed experiments \ref{deta} on various types of image disturbance amplitudes.

\subsection{Robustness evaluation against VAE-based attacks and diffusion based attacks.} In Table \ref{tab:deeplearningattack}, we evaluated the accuracy of various watermarking schemes when faced with deep learning-based attacks. The experiments consist of VAE-based attack and diffision-based attack which is the latest watermark attack. The table reveals that the majority of schemes are unable to withstand this type of reconstruction attack. From the results, it is evident that DiffuseTrace exhibits a significant advantage in countering VAE-based attacks utilizing diffusion models.
SSL Watermarks and Stable Signatures all exhibit low watermark detection rates, indicating that they are unable to resist both VAE-based and diffusion based attacks. For diffusion based attacks, except for DiffuseTrace, other watermark schemes have experienced a significant drop in bit accuracy. In subsequent experiments, we increased the intensity of the diffusion attack. From the results, it is evident that DiffuseTrace exhibits significantly higher resilience against reconstruction attacks compared to other methods. Even the VAE-attack with a quality coefficient of 1 or the diffusion-based attack with 150 denoise steps do not fundamentally affect the stability of the watermark and only the DiffuseTrace was able to maintain accuracy in high-intensity reconstruction. Reconstruction attacks are achieved by maintaining semantic accuracy, continuously adding noise to destroy watermarks and continuously reconstructing and restoring images to obtain images without watermarks.  However, this process essentially does not alter the semantic consistency of the image nor does it significantly alter the initial latent variables of image inversion. Therefore, DiffuseTrace can remain stable under reconstruction attacks. 

\begin{table}[t]
\caption{Bit Accuracy/Detection Accuracy Under Generative Attack}
\centering
\setlength{\tabcolsep}{1.3mm}{
\begin{tabular}{*{5}{c}}
\toprule
\multicolumn{2}{c}{\multirow{2}{*}{\textbf{Method}}} & \multicolumn{2}{c}{\textbf{VAE A.}} & \textbf{Diffusion A.} \\
\cmidrule(lr){3-4} \cmidrule(l){5-5}
\multicolumn{2}{c}{} & \textbf{Bmshj18} \cite{balle2018variational} & \textbf{Cheng20} \cite{cheng2020learned} & \textbf{Zhao23} \cite{zhao2023generative} \\
\midrule
\multirow{2}{*}{\begin{tabular}[c]{@{}c@{}}Traditional \\ Wm.\end{tabular}}
&DwtDct&0.524/0.000&0.517/0.012&0.489/0.000\\
&D.Svd&0.504/0.000&0.512/0.013&0.523/0.014\\
\midrule
\multirow{4}{*}{\begin{tabular}[c]{@{}c@{}}Enc.-Dec.\\ Wm.\end{tabular}}
&RivaGan&0.611/0.063&0.632/0.070&0.588/0.070\\
&Hidden&0.621/0.170&0.641/0.198&0.497/0.009\\
&S.Stamp&0.979/0.999&0.965/0.999&0.852/0.927\\
&SSL Wm.&0.623/0.123&0.631/0.144&0.655/0.149\\
\midrule
\multirow{2}{*}{\begin{tabular}[c]{@{}c@{}}VAE-based \\ Wm.\end{tabular}}
&\multirow{2}{*}{\begin{tabular}[c]{@{}c@{}}Stable \\ Signature\end{tabular}}&\multirow{2}{*}{0.616/0.224}&\multirow{2}{*}{0.682/0.409}&\multirow{2}{*}{0.541/0.014}\\
\\
\midrule
\multirow{2}{*}{\begin{tabular}[c]{@{}c@{}}Latent-based \\ Wm.\end{tabular}}
&Tree-ring&------ /0.993&------ /0.999&------ /0.994\\
&\textbf{Ours(ori)}&0.972/\textbf{0.999}&0.967/\textbf{0.999}&0.970/\textbf{0.999}\\
\bottomrule
\end{tabular}}
\label{tab:deeplearningattack}
\end{table}

\subsection{Ablation Experiments}
In this section, we experimentally quantify the impact of several key hyperparameters mentioned in the theoretical analysis \ref{sec:theory analysis} on the inaccuracy. We consider the impact of the guidance scale used during the generation phase, the inference steps employed during the inversion phase, and the version of the model on watermark detection in order to demonstrate the effectiveness of DiffuseTrace.

\textbf{Ablation on Guidance Scale.} In the theoretical analysis \ref{subsec:offset}, we elaborate on the reasons why the guidance scale introduces errors into the experiments. In the following experiments, we quantify the impact of the guidance scale on the DiffuseTrace watermarking scheme through experimentation. For the ablation experiment on the guidance scale, the scheduler is set the dpm++ \cite{lu2022dpm} scheduler. The experimental setup includes setting both the inference steps and reverse inference steps to 20. We adjust the guidance scale to assess its influence on the experimental results.

\begin{figure}[t]
    \centering
    \subfloat[Guidance Scale]{ %
        \includegraphics[width=0.49\linewidth]{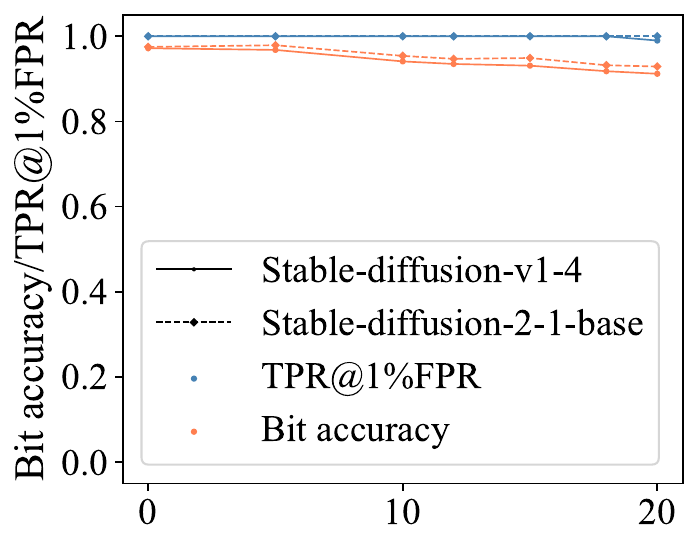}
        \label{subfig:guidance}
    }
    \hspace{-1.3em}
    \subfloat[Reverse Steps]{ %
        \includegraphics[width=0.49\linewidth]{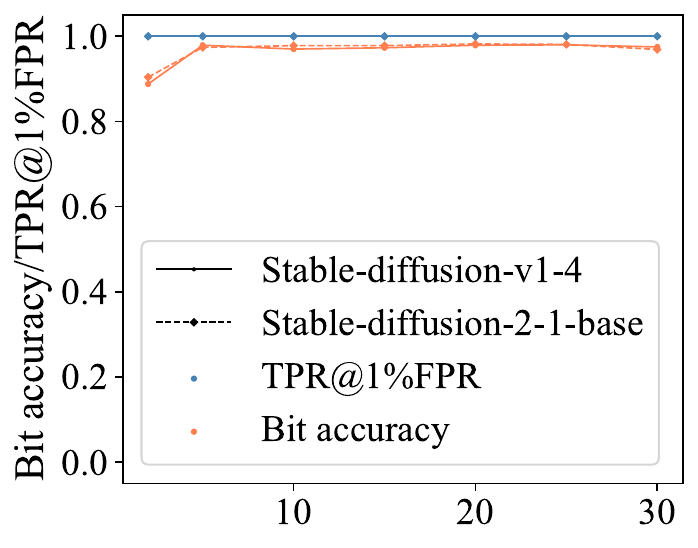}
        \label{subfig:reverse}
    }
    \caption{The figure (a) illustrates the ablation experiment concerning the guidance scale, where adjusting the guidance scale leads to a gradual decrease in the watermark's bit accuracy, while the watermark detection rate remains stable. The figure (b) shows the results of the ablation study on reverse inference steps, where the bit rate detected stabilizes after two inference steps.}
    \label{exp:ablation exp}
\end{figure}

The specific experimental results depicted in the graph \ref{exp:ablation exp} show that as the guidance scale increases during the inference stage, the bit accuracy gradually decreases, while the detection accuracy remains relatively stable within the guidance scale range of 0 to 20. This indicates that the watermark detection accuracy is maintained, demonstrating the robustness of the watermark. Thus, users can freely adjust the guidance scale during the image generation stage while still ensuring traceability of the watermark. Deploying the diffusion model as a service can provide users with the option to adjust the guidance scale hyperparameter, which will not significantly affect watermark detection.

\textbf{Ablation on Inference Steps.}
The ablation experiment for reverse inference steps employed the DPM++ scheduler, with an inference step setting of 20 and a guidance scale set to 5. The evaluation of the experiment's results involves adjusting the reverse inference steps to assess their impact.

The experimental results, as depicted in the figure \ref{exp:ablation exp}, indicate that after 5 inference steps of inversion, the watermark detection rate stabilizes. This suggests that the number of inference steps does not significantly affect the accuracy of detection. Therefore, during the detection phase, to increase efficiency, a small number of reverse inference steps can be employed to extract the image watermark.

%% file: conclusion.tex
\section{Conclusion}
\label{sec:conclusion}
In this paper we propose DiffuseTrace, a plug-in multibit watermarking module to protect copyright of diffusion models and trace generated images sematically.  
DiffuseTrace extends semantic image watermarking of latent diffusion models further into multi-bit scenarios. DiffuseTrace does not rely on the balance between image quality and watermark robustness and has significant advantages in image quality compared to previous watermarking schemes. Compared to state-of-the-art schemes, DiffuseTrace demonstrates prominent performance  against variational autoencoders and diffusion-based watermark attacks.
Due to its flexibility and generalizability, DiffuseTrace can be seamlessly applied to copyright protection in diffusion models, recognition of machine-generated images and user traceability in machine-generated image services. 
We assess the security of the watermarking scheme through theoretical analysis and demonstrate its robustness against image processing attacks and state-of-the-art image watermarking attack schemes through experiments.

%% file: main.bbl
\begin{thebibliography}{10}
\providecommand{\url}[1]{#1}
\csname url@samestyle\endcsname
\providecommand{\newblock}{\relax}
\providecommand{\bibinfo}[2]{#2}
\providecommand{\BIBentrySTDinterwordspacing}{\spaceskip=0pt\relax}
\providecommand{\BIBentryALTinterwordstretchfactor}{4}
\providecommand{\BIBentryALTinterwordspacing}{\spaceskip=\fontdimen2\font plus
\BIBentryALTinterwordstretchfactor\fontdimen3\font minus \fontdimen4\font\relax}
\providecommand{\BIBforeignlanguage}[2]{{%
\expandafter\ifx\csname l@#1\endcsname\relax
\typeout{** WARNING: IEEEtran.bst: No hyphenation pattern has been}%
\typeout{** loaded for the language `#1'. Using the pattern for}%
\typeout{** the default language instead.}%
\else
\language=\csname l@#1\endcsname
\fi
#2}}
\providecommand{\BIBdecl}{\relax}
\BIBdecl

\bibitem{ho2020denoising}
J.~Ho, A.~Jain, and P.~Abbeel, ``Denoising diffusion probabilistic models,'' \emph{Advances in Neural Information Processing Systems}, vol.~33, pp. 6840--6851, 2020.

\bibitem{song2020score}
Y.~Song, J.~Sohl-Dickstein, D.~P. Kingma, A.~Kumar, S.~Ermon, and B.~Poole, ``Score-based generative modeling through stochastic differential equations,'' \emph{arXiv preprint arXiv:2011.13456}, 2020.

\bibitem{dhariwal2021diffusion}
P.~Dhariwal and A.~Nichol, ``Diffusion models beat gans on image synthesis,'' \emph{Advances in Neural Information Processing Systems}, vol.~34, pp. 8780--8794, 2021.

\bibitem{rombach2022high}
R.~Rombach, A.~Blattmann, D.~Lorenz, P.~Esser, and B.~Ommer, ``High-resolution image synthesis with latent diffusion models,'' in \emph{Proceedings of the IEEE/CVF Conference on Computer Vision and Pattern Recognition}, 2022, pp. 10\,684--10\,695.

\bibitem{saharia2022photorealistic}
C.~Saharia, W.~Chan, S.~Saxena, L.~Li, J.~Whang, E.~L. Denton, K.~Ghasemipour, R.~Gontijo~Lopes, B.~Karagol~Ayan, T.~Salimans \emph{et~al.}, ``Photorealistic text-to-image diffusion models with deep language understanding,'' \emph{Advances in Neural Information Processing Systems}, vol.~35, pp. 36\,479--36\,494, 2022.

\bibitem{zhang2023adding}
L.~Zhang, A.~Rao, and M.~Agrawala, ``Adding conditional control to text-to-image diffusion models,'' in \emph{Proceedings of the IEEE/CVF International Conference on Computer Vision}, 2023, pp. 3836--3847.

\bibitem{nichol2021glide}
A.~Nichol, P.~Dhariwal, A.~Ramesh, P.~Shyam, P.~Mishkin, B.~McGrew, I.~Sutskever, and M.~Chen, ``Glide: Towards photorealistic image generation and editing with text-guided diffusion models,'' \emph{arXiv preprint arXiv:2112.10741}, 2021.

\bibitem{brooks2023instructpix2pix}
T.~Brooks, A.~Holynski, and A.~A. Efros, ``Instructpix2pix: Learning to follow image editing instructions,'' in \emph{Proceedings of the IEEE/CVF Conference on Computer Vision and Pattern Recognition}, 2023, pp. 18\,392--18\,402.

\bibitem{saharia2022palette}
C.~Saharia, W.~Chan, H.~Chang, C.~Lee, J.~Ho, T.~Salimans, D.~Fleet, and M.~Norouzi, ``Palette: Image-to-image diffusion models,'' in \emph{ACM SIGGRAPH 2022 Conference Proceedings}, 2022, pp. 1--10.

\bibitem{lugmayr2022repaint}
A.~Lugmayr, M.~Danelljan, A.~Romero, F.~Yu, R.~Timofte, and L.~Van~Gool, ``Repaint: Inpainting using denoising diffusion probabilistic models,'' in \emph{Proceedings of the IEEE/CVF Conference on Computer Vision and Pattern Recognition}, 2022, pp. 11\,461--11\,471.

\bibitem{saharia2022image}
C.~Saharia, J.~Ho, W.~Chan, T.~Salimans, D.~J. Fleet, and M.~Norouzi, ``Image super-resolution via iterative refinement,'' \emph{IEEE Transactions on Pattern Analysis and Machine Intelligence}, vol.~45, no.~4, pp. 4713--4726, 2022.

\bibitem{esser2021taming}
P.~Esser, R.~Rombach, and B.~Ommer, ``Taming transformers for high-resolution image synthesis,'' in \emph{Proceedings of the IEEE/CVF Conference on Computer Vision and Pattern Recognition}, 2021, pp. 12\,873--12\,883.

\bibitem{ramesh2021zero}
A.~Ramesh, M.~Pavlov, G.~Goh, S.~Gray, C.~Voss, A.~Radford, M.~Chen, and I.~Sutskever, ``Zero-shot text-to-image generation,'' in \emph{International Conference on Machine Learning}.\hskip 1em plus 0.5em minus 0.4em\relax Pmlr, 2021, pp. 8821--8831.

\bibitem{ramesh2022hierarchical}
A.~Ramesh, P.~Dhariwal, A.~Nichol, C.~Chu, and M.~Chen, ``Hierarchical text-conditional image generation with clip latents,'' \emph{arXiv preprint arXiv:2204.06125}, vol.~1, no.~2, p.~3, 2022.

\bibitem{blattmann2023align}
A.~Blattmann, R.~Rombach, H.~Ling, T.~Dockhorn, S.~W. Kim, S.~Fidler, and K.~Kreis, ``Align your latents: High-resolution video synthesis with latent diffusion models,'' in \emph{Proceedings of the IEEE/CVF Conference on Computer Vision and Pattern Recognition}, 2023, pp. 22\,563--22\,575.

\bibitem{ho2022imagen}
J.~Ho, W.~Chan, C.~Saharia, J.~Whang, R.~Gao, A.~Gritsenko, D.~P. Kingma, B.~Poole, M.~Norouzi, D.~J. Fleet \emph{et~al.}, ``Imagen video: High definition video generation with diffusion models,'' \emph{arXiv preprint arXiv:2210.02303}, 2022.

\bibitem{cox2007digital}
I.~Cox, M.~Miller, J.~Bloom, J.~Fridrich, and T.~Kalker, \emph{Digital watermarking and steganography}.\hskip 1em plus 0.5em minus 0.4em\relax Morgan kaufmann, 2007.

\bibitem{ronneberger2015u}
O.~Ronneberger, P.~Fischer, and T.~Brox, ``U-net: Convolutional networks for biomedical image segmentation,'' in \emph{Medical Image Computing and Computer-assisted Intervention--MICCAI 2015: 18th international conference, Munich, Germany, October 5-9, 2015, proceedings, part III 18}.\hskip 1em plus 0.5em minus 0.4em\relax Springer, 2015, pp. 234--241.

\bibitem{zhao2023invisible}
X.~Zhao, K.~Zhang, Z.~Su, S.~Vasan, I.~Grishchenko, C.~Kruegel, G.~Vigna, Y.~Wang, and L.~Li, ``Invisible image watermarks are provably removable using generative ai,'' \emph{Saastha Vasan, Ilya Grishchenko, Christopher Kruegel, Giovanni Vigna, Yu-Xiang Wang, and Lei Li,“Invisible image watermarks are provably removable using generative ai,” Aug}, 2023.

\bibitem{wang2020cnn}
S.-Y. Wang, O.~Wang, R.~Zhang, A.~Owens, and A.~A. Efros, ``Cnn-generated images are surprisingly easy to spot... for now,'' in \emph{Proceedings of the IEEE/CVF Conference on Computer vision and pattern recognition}, 2020, pp. 8695--8704.

\bibitem{frank2020leveraging}
J.~Frank, T.~Eisenhofer, L.~Sch{\"o}nherr, A.~Fischer, D.~Kolossa, and T.~Holz, ``Leveraging frequency analysis for deep fake image recognition,'' in \emph{International conference on Machine Learning}.\hskip 1em plus 0.5em minus 0.4em\relax PMLR, 2020, pp. 3247--3258.

\bibitem{tan2023learning}
C.~Tan, Y.~Zhao, S.~Wei, G.~Gu, and Y.~Wei, ``Learning on gradients: Generalized artifacts representation for gan-generated images detection,'' in \emph{Proceedings of the IEEE/CVF Conference on Computer Vision and Pattern Recognition}, 2023, pp. 12\,105--12\,114.

\bibitem{wang2023dire}
Z.~Wang, J.~Bao, W.~Zhou, W.~Wang, H.~Hu, H.~Chen, and H.~Li, ``Dire for diffusion-generated image detection,'' in \emph{Proceedings of the IEEE/CVF International Conference on Computer Vision}, 2023, pp. 22\,445--22\,455.

\bibitem{al2007combined}
A.~Al-Haj, ``Combined dwt-dct digital image watermarking,'' \emph{Journal of Computer Science}, vol.~3, no.~9, pp. 740--746, 2007.

\bibitem{liu2019optimized}
J.~Liu, J.~Huang, Y.~Luo, L.~Cao, S.~Yang, D.~Wei, and R.~Zhou, ``An optimized image watermarking method based on hd and svd in dwt domain,'' \emph{IEEE Access}, vol.~7, pp. 80\,849--80\,860, 2019.

\bibitem{zhu2018hidden}
J.~Zhu, R.~Kaplan, J.~Johnson, and L.~Fei-Fei, ``Hidden: Hiding data with deep networks,'' in \emph{Proceedings of the European Conference on Computer Vision}, 2018, pp. 657--672.

\bibitem{tancik2020stegastamp}
M.~Tancik, B.~Mildenhall, and R.~Ng, ``Stegastamp: Invisible hyperlinks in physical photographs,'' in \emph{Proceedings of the IEEE/CVF Conference on Computer Vision and Pattern Recognition}, 2020, pp. 2117--2126.

\bibitem{zhao2023recipe}
Y.~Zhao, T.~Pang, C.~Du, X.~Yang, N.-M. Cheung, and M.~Lin, ``A recipe for watermarking diffusion models,'' \emph{arXiv preprint arXiv:2303.10137}, 2023.

\bibitem{fernandez2023stable}
P.~Fernandez, G.~Couairon, H.~J{\'e}gou, M.~Douze, and T.~Furon, ``The stable signature: Rooting watermarks in latent diffusion models,'' in \emph{Proceedings of the IEEE/CVF International Conference on Computer Vision}, 2023, pp. 22\,466--22\,477.

\bibitem{xiong2023flexible}
C.~Xiong, C.~Qin, G.~Feng, and X.~Zhang, ``Flexible and secure watermarking for latent diffusion model,'' in \emph{Proceedings of the 31st ACM International Conference on Multimedia}, 2023, pp. 1668--1676.

\bibitem{wen2023tree}
Y.~Wen, J.~Kirchenbauer, J.~Geiping, and T.~Goldstein, ``Tree-ring watermarks: Fingerprints for diffusion images that are invisible and robust,'' \emph{arXiv preprint arXiv:2305.20030}, 2023.

\bibitem{zhang2024robust}
L.~Zhang, X.~Liu, A.~V. Martin, C.~X. Bearfield, Y.~Brun, and H.~Guan, ``Robust image watermarking using stable diffusion,'' \emph{arXiv preprint arXiv:2401.04247}, 2024.

\bibitem{dabov2007image}
K.~Dabov, A.~Foi, V.~Katkovnik, and K.~Egiazarian, ``Image denoising by sparse 3-d transform-domain collaborative filtering,'' \emph{IEEE Transactions on Image Processing}, vol.~16, no.~8, pp. 2080--2095, 2007.

\bibitem{balle2018variational}
J.~Ball{\'e}, D.~Minnen, S.~Singh, S.~J. Hwang, and N.~Johnston, ``Variational image compression with a scale hyperprior,'' \emph{arXiv preprint arXiv:1802.01436}, 2018.

\bibitem{cheng2020learned}
Z.~Cheng, H.~Sun, M.~Takeuchi, and J.~Katto, ``Learned image compression with discretized gaussian mixture likelihoods and attention modules,'' in \emph{Proceedings of the IEEE/CVF Conference on Computer Vision and Pattern Recognition}, 2020, pp. 7939--7948.

\bibitem{zhao2023generative}
X.~Zhao, K.~Zhang, Y.-X. Wang, and L.~Li, ``Generative autoencoders as watermark attackers: Analyses of vulnerabilities and threats,'' \emph{arXiv preprint arXiv:2306.01953}, 2023.

\bibitem{cemgil2020autoencoding}
T.~Cemgil, S.~Ghaisas, K.~Dvijotham, S.~Gowal, and P.~Kohli, ``The autoencoding variational autoencoder,'' \emph{Advances in Neural Information Processing Systems}, vol.~33, pp. 15\,077--15\,087, 2020.

\bibitem{song2020denoising}
J.~Song, C.~Meng, and S.~Ermon, ``Denoising diffusion implicit models,'' \emph{arXiv preprint arXiv:2010.02502}, 2020.

\bibitem{lu2022dpm}
C.~Lu, Y.~Zhou, F.~Bao, J.~Chen, C.~Li, and J.~Zhu, ``Dpm-solver++: Fast solver for guided sampling of diffusion probabilistic models,'' \emph{arXiv preprint arXiv:2211.01095}, 2022.

\bibitem{doersch2016tutorial}
C.~Doersch, ``Tutorial on variational autoencoders,'' \emph{arXiv preprint arXiv:1606.05908}, 2016.

\bibitem{mokady2023null}
R.~Mokady, A.~Hertz, K.~Aberman, Y.~Pritch, and D.~Cohen-Or, ``Null-text inversion for editing real images using guided diffusion models,'' in \emph{Proceedings of the IEEE/CVF Conference on Computer Vision and Pattern Recognition}, 2023, pp. 6038--6047.

\bibitem{hong2024exact}
S.~Hong, K.~Lee, S.~Y. Jeon, H.~Bae, and S.~Y. Chun, ``On exact inversion of dpm-solvers,'' in \emph{Proceedings of the IEEE/CVF Conference on Computer Vision and Pattern Recognition}, 2024, pp. 7069--7078.

\bibitem{zhang2018unreasonable}
R.~Zhang, P.~Isola, A.~A. Efros, E.~Shechtman, and O.~Wang, ``The unreasonable effectiveness of deep features as a perceptual metric,'' in \emph{Proceedings of the IEEE Conference on Computer Vision and Pattern Recognition}, 2018, pp. 586--595.

\bibitem{lin2014microsoft}
T.-Y. Lin, M.~Maire, S.~Belongie, J.~Hays, P.~Perona, D.~Ramanan, P.~Doll{\'a}r, and C.~L. Zitnick, ``Microsoft coco: Common objects in context,'' in \emph{Computer Vision--ECCV 2014: 13th European Conference, Zurich, Switzerland, September 6-12, 2014, Proceedings, Part V 13}.\hskip 1em plus 0.5em minus 0.4em\relax Springer, 2014, pp. 740--755.

\bibitem{wang2022diffusiondb}
Z.~J. Wang, E.~Montoya, D.~Munechika, H.~Yang, B.~Hoover, and D.~H. Chau, ``Diffusiondb: A large-scale prompt gallery dataset for text-to-image generative models,'' \emph{arXiv preprint arXiv:2210.14896}, 2022.

\bibitem{zhang2019robust}
K.~A. Zhang, L.~Xu, A.~Cuesta-Infante, and K.~Veeramachaneni, ``Robust invisible video watermarking with attention,'' \emph{arXiv preprint arXiv:1909.01285}, 2019.

\bibitem{fernandez2022watermarking}
P.~Fernandez, A.~Sablayrolles, T.~Furon, H.~J{\'e}gou, and M.~Douze, ``Watermarking images in self-supervised latent spaces,'' in \emph{ICASSP 2022-2022 IEEE International Conference on Acoustics, Speech and Signal Processing}.\hskip 1em plus 0.5em minus 0.4em\relax IEEE, 2022, pp. 3054--3058.

\bibitem{Radford2021LearningTV}
A.~Radford, J.~W. Kim, C.~Hallacy, A.~Ramesh, G.~Goh, S.~Agarwal, G.~Sastry, A.~Askell, P.~Mishkin, J.~Clark, G.~Krueger, and I.~Sutskever, ``Learning transferable visual models from natural language supervision,'' in \emph{ICML}, 2021.

\bibitem{mittal2012making}
A.~Mittal, R.~Soundararajan, and A.~C. Bovik, ``Making a “completely blind” image quality analyzer,'' \emph{IEEE Signal Processing Letters}, vol.~20, no.~3, pp. 209--212, 2012.

\bibitem{venkatanath2015blind}
N.~Venkatanath, D.~Praneeth, M.~C. Bh, S.~S. Channappayya, and S.~S. Medasani, ``Blind image quality evaluation using perception based features,'' in \emph{2015 Twenty First National Conference on Communications}.\hskip 1em plus 0.5em minus 0.4em\relax IEEE, 2015, pp. 1--6.

\bibitem{Rombach_2022_CVPR}
R.~Rombach, A.~Blattmann, D.~Lorenz, P.~Esser, and B.~Ommer, ``High-resolution image synthesis with latent diffusion models,'' in \emph{Proceedings of the IEEE/CVF Conference on Computer Vision and Pattern Recognition (CVPR)}, June 2022, pp. 10\,684--10\,695.

\end{thebibliography}
